\documentclass[a4paper,
twocolumn,preprintnumbers,citeautoscript,newabstract,aps,nofootinbib,superscriptaddress]{revtex4-2}
\usepackage{amsmath}
\usepackage{amsfonts}
\usepackage{dsfont}
\usepackage{amssymb}
\usepackage{slashed}
\usepackage[hidelinks=true]{hyperref}
\usepackage{graphicx}
\usepackage{color}
\usepackage{mathtools}

\let\oldFootnote\footnote
\newcommand\nextToken\relax

\renewcommand\footnote[1]{%
    \oldFootnote{#1}\futurelet\nextToken\isFootnote}

\newcommand\isFootnote{%
    \ifx\footnote\nextToken\textsuperscript{,}\fi}

\newcommand{\U}[1]{\ensuremath{\mathrm{U}(#1)}}
\newcommand{\Z}[1]{\ensuremath{\mathds{Z}_{#1}}}
\newcommand{\I}{\ensuremath{\mathrm{i}}}
\newcommand{\e}{\ensuremath{\mathrm{e}}}

\newcommand{\Out}[1]{\ensuremath{\mathrm{Out}(#1)}}
\newcommand{\rep}[2][]{\ensuremath{\boldsymbol{#2}#1}}

\newtheorem{theorem}{Finding}
\newtheorem{corollary}{Corollary}

\begin{document}
\title{Outer automorphisms are sufficient conditions for RG fixed points}

\author{\vspace{-0.1cm} Thede de Boer}
\affiliation{Max-Planck-Institut f\"ur Kernphysik, Saupfercheckweg 1, 69117 Heidelberg, Germany}
\author{Andreas Trautner}\email[]{trautner@cftp.ist.utl.pt}
\affiliation{CFTP, Departamento de F\'isica, Instituto Superior T\'ecnico, Universidade de Lisboa, Avenida Rovisco Pais 1, 1049 Lisboa, Portugal}

\begin{abstract}\noindent
We point out that the existence of an outer automorphism (Out) is a sufficient condition for the existence of a fixed hyperplane (fixed point, separatrix) in the renormalization group~(RG) flow of a Quantum Field Theory~(QFT). The corresponding RG fixed hyperplane is determined by a symmetry argument and can be computed without resorting to perturbation theory. This provides the mathematical underpinning of 't\,Hooft's technical naturalness argument, and results in a systematic way to derive non-perturbative all-order constraints on the RG beta functions. If an Out exists, the symmetry of the fully coupled system of beta functions is larger than the symmetry of the action. We also stress the importance of including goofy transformations in these considerations.
\end{abstract}
\maketitle

\enlargethispage{2.6cm}
\section{Introduction}
Symmetries play a fundamental role in all aspects of physics. In QFTs, the partition function is typically constrained by imposing symmetries and assigning the quantum fields to representations of these. This is used as a classifying principle and constrains the interactions of particles in a non-perturbative way.

In addition to the symmetries imposed in its construction, a QFT may have additional \textit{possible} or \textit{prospective} symmetries. For a fixed field content, \textit{prospective} symmetries are only realized for specific values of the coupling coefficients of the QFT. Finding possible symmetries of this kind is, in general, a non-trivial task. 

It helps to view such \textit{prospective} symmetries as being explicitly broken at a generic point in the parameter space of the original QFT. On certain hyperplanes in the parameter space of the theory, though, the \textit{prospective} symmetries become exact. The RG flow will not cross such hyperplanes, implying that they form (partial) RG fixed points (attractive, repulsive or separatrices). A brute force way to spot possible symmetries of this kind, hence, is to compute quantum corrections in the QFT to set up the associated Gell-Mann-Low RG beta functions in perturbation theory. The solutions to the coupled system of $\beta$ functions then are expected to reveal all the fixed hyperplanes which correspond to the preserved \textit{prospective} symmetries. This is a commonly followed route, see e.g.~the recent~\cite{Berezhiani:2024rth} and references therein, where the \textit{prospective} symmetries that arise as boundaries of the RG flow are referred to as \textit{emergent} symmetries. Symmetries under discussion here include discrete and continuous internal and space-time symmetries -- as we will use as examples in this manuscript -- and the general logic also applies to supersymmetry, gauge symmetry (gauge redundancies), and conformal and/or scale symmetry.

The fact that the RG evolution of theories does not cross or run away from symmetry enhanced points is a folk wisdom that goes back to the original technical naturalness argument of 't\,Hooft~\cite{tHooft:1979rat}. He argued that the $\beta$ function of a symmetry-breaking parameter has to be proportional to the symmetry-breaking parameter itself, such that if the breaking is set to zero also the corresponding $\beta$ function vanishes. 

\newpage
Our work constitutes a formal generalization and extension of 't\,Hooft's argument. We point out that at least some \textit{prospective} symmetries of QFTs are, in fact, given by the \textit{outer automorphisms} (Outs) of those QFTs. Relevant Outs can be computed directly from the symmetry \textit{and} representation content of a QFT.\footnote{%
\label{fot:maximalbreaking}%
The representation content is important because there can also exist Outs which map fields in irreps to irreps which are not included in the original QFT. Such transformations are called explicitly and  \textit{maximally} broken. The according transformations cannot be restored by choosing parameter values; two famous examples are the parity (P) and charge-conjugation (C) transformations in the Standard Model.}
This allows for a much easier route to compute the associated RG fixed points than brute force computing and numerically solving the $\beta$ functions. This simplicity arises because the sheer existence of each Out gives rise to a non-perturbative constraint on the fully coupled system of $\beta$ functions. Crucially, the Out constraint is fulfilled independently of whether the corresponding symmetry is realized in the QFT.
This implies that the system of $\beta$ functions generally has symmetries that are not symmetries of the action. In fact, the existence of such constraints is the mathematical reason for the existence of RG fixed points in the first place.

Our main result is that the existence of a consistent (not maximally broken) \textit{outer automorphism}~(Out) of a QFT provides a sufficient condition for the existence of an RG fixed hyperplane. In this way, the \textit{prospective} or \textit{emergent} symmetries can be computed in a symmetry based way without resorting to perturbation theory. Our argument can be applied to any QFT. The argument is general, in the sense that it applies to parameters of any mass dimension (couplings, masses, wave function renormalization~(WFR) coefficients, \textit{etc.}). With some care, our argument can also be applied to non-renormalizable effective theories.

We first present our argument intuitively, then give two simple examples. Afterwards we outline the formal derivation on general grounds.
Finally we give some comments and an outlook after which we conclude.

\section{General Argument (in Words)}\enlargethispage{1.2cm}
Mathematically, symmetries are described by groups. Automorphisms of groups map group elements among themselves while leaving invariant the group's algebra. Some automorphisms are generated by the group elements themselves, e.g.\ $g\mapsto h \cdot g \cdot h^{-1}$ for $h,g\in G$ where $G$ is the symmetry group. These automorphisms are called \textit{inner}. Any automorphism that is not \textit{inner} is called an \textit{outer automorphism}~(Out).\footnote{%
For an introduction to Outs and their action on QFTs, see~\cite{Fallbacher:2015rea, Trautner:2016ezn, Trautner:2017vlz,Fallbacher:2015upf}.} While inner automorphisms, by definition, act trivially on the spectrum of irreps of a group, an Out generally acts as a non-trivial permutation on the irreps of a group.

Inner automorphisms of the preserved symmetry groups of a QFT leave the theory invariant by construction.\footnote{This is true at the classical level and can be modified by quantum corrections in the case of anomalous ``symmetries'' which are not actual symmetries in this sense.} By contrast, outer automorphisms (Outs) are not symmetries of the original theory. Nonetheless, Outs act on the symmetry and its representations and, hence, can also be used to act on a QFT.

Here it is important to distinguish Outs that map all irreps included in a QFT among themselves, from Outs that would map irreps of a QFT to irreps which are not originally present in the spectrum of the theory. In the latter case, the Out can never be preserved within the original QFT, and the according symmetry is called \textit{maximally} broken~(see footnote~\ref{fot:maximalbreaking}). We will exclude such transformation from the following discussion. By contrast, in the first case the QFT and, in particular, the spectrum of contained irreps is mapped to itself by the action of an Out. The action of such an Out then can \textit{equivalently} be described as a mapping in the parameter space of a theory without changing the basis of operators.\footnote{
This point is crucial. Not changing the operator basis of the theory implies that the functional form of the all-order $\beta$ functions does not change. This is not the case for arbitrary transformations on the fields.}\footnote{
A well known comprehensible example for this case is the CP transformation of the Standard Model. Instead of acting on the space-time coordinate and all fields of the theory as well defined Out (see e.g.~\cite{Grimus:1995zi, Trautner:2016ezn}), this transformation can \textit{equivalently} be described by the parameter space mapping $\delta_{\mathrm{CKM}}\mapsto-\delta_{\mathrm{CKM}}$. The existence of the corresponding Out restricts $\beta_{\delta_{\mathrm{CKM}}}\propto\delta_{\mathrm{CKM}}$, implying that values of $\delta_{\mathrm{CKM}}=0$ are RG fixed points corresponding to the Out-enhanced symmetry.}
That is, couplings of a theory generally transform covariantly under the action of an Out and can, therefore, be combined into irreps under the Outs~\cite{Fallbacher:2015rea}. 

Consider now the fully coupled system of $\beta$ functions ($\lambda_{n=1,2,\dots}$ here should be thought of as running over all couplings, masses, WFR coefficients \textit{etc.}), 
\begin{equation}\label{eq:beta}
\beta_{\lambda_n}~\equiv~\mu\frac{\mathrm{d}\,\lambda_n}{\mathrm{d}\mu}~=~f_{n}(\lambda_1,\lambda_2,\dots)\;.
\end{equation}
By definition, the l.h.s.\ is linear in the couplings (differentiation is a linear operator). The covariant transformation behavior of couplings under an Out, therefore, directly translates to the necessarily covariant transformation behavior of the non-linear $\beta$ functions on the r.h.s.\; 
Writing the mapping on the couplings as $\lambda_n\mapsto F_n(\lambda)$, we will in Sec.~\ref{sec:GenArg2} derive that the $\beta$ functions must obey\footnote{To clarify functional dependencies in this paper we use $h(\lambda)$ to denote dependence on \textit{all} $\lambda_n$'s and we use $h(\lambda_k)$ and $h(F_k(\lambda))$, which likewise depend on all $\lambda_n$'s, when it is necessary to specify which $\lambda_n$ or $F_n(\lambda)$ goes into the $k$-th slot of a function.}
\begin{equation}\label{eq:beta_trafo}
\beta_{\lambda_n}(F_k(\lambda))=\sum_m\frac{\partial\, F_n(\lambda)}{\partial\lambda_m}\beta_{\lambda_m}(\lambda_k).
\end{equation}
The Out, therefore, acts as a symmetry\footnote{
By symmetry of a system of differential equations we mean a transformation that maps one solution of the system to another solution, see e.g.~\cite[Def.~2.23]{Olver:1986}.}
of the fully coupled system of $\beta$ functions of a theory.

This implies that the mere existence of an Out gives rise to the non-perturbative, all order exact constraint \eqref{eq:beta_trafo} on the $\beta$ functions. Since the $\beta$ functions have to transform covariantly, their respective functional form has to be spanned by combination of couplings that likewise transform covariantly \textit{and in the same irrep} as the $\beta$ functions themselves. For simple algebraic reasons, the number of possible covariant coupling structures in the $\beta$ functions terminates at typically rather low orders such that the structure of the $\beta$ functions can fully be determined at the non-perturbative level. 
The larger the number of possible independent Outs, the more restrictive are the combined overall constraints on the $\beta$ functions. This can lead to situations where the structural functional form of the $\beta$ functions in terms of non-trivially transforming covariant combination of couplings is completely determined to all orders. Other, higher-order corrections can only enter those functions in the form of Out- and symmetry invariant singlet prefactors.

Requiring an Out to join the actual symmetries of a theory corresponds to demanding that all non-trivially transforming covariant combination of couplings must vanish. As a direct consequence, this would lead to the vanishing of all non-trivially transforming $\beta$ functions, thereby reducing the rank of the system of $\beta$ functions. This establishes the (partial) RG fixed point to all orders. This is our more general formulation of 't\,Hooft's original argument. It extends the usual intuition in the sense that, (i) it applies also far away from the actual symmetry enhanced points, and (ii) it potentially includes more than one non-trivially transforming covariant.

Note that the constraints on the $\beta$ functions neither require that a possible Out is imposed, nor that we are perturbatively close to the limit where an Out would become a conserved symmetry. The fully coupled system of $\beta$ functions is symmetric under all \textit{all possible} Outs in any case, even if we are parametrically far away from points that realize the symmetry.

We stress that our arguments apply to any kind of active transformation on the quantum fields that can equivalently be described as a $1:1$ mapping on the parameter space of a QFT. Since this implies the invariance (as a set) of the already present and symmetry constrained operators we think such a map should always correspond to an Out.\footnote{
In addition to symmetry extensions by Outs, Ref.~\cite{Doring:2024kdg} discusses the second (exhaustive) possibility of so-called ``unorthodox'' extensions. We conjecture that unorthodox extensions do \textit{not} correspond to $1:1$ mappings in the parameter space of a QFT but impose non-invertible constraints on the couplings of a theory.}

A special case is that there are no symmetries in the original QFT to start with. The set of possible Outs (of the then trivial group $G$) then contains all possible basis transformations. Just like Outs of any other symmetry, the basis changes of an entirely unconstrained theory induce linear transformations on the parameter space of the theory under which couplings and their associated $\beta$ functions transform covariantly.\footnote{
It is known to practitioners that $\beta$ functions in general models can be organized to all orders in basis-change covariants, see e.g.~\cite{Bednyakov:2018cmx, Bednyakov:2025sri}. Our argument is more general than that, as it includes all Outs, noting that basis changes are trivial Outs only in the case that there is no other symmetry constraining the potential.}

Beyond formalizing 't\,Hooft's original argument, very importantly our discussion also applies for the recently discovered \textit{goofy} transformations~\cite{Ferreira:2023dke,Trautner:2025yxz}. Goofy trafos also correspond to Outs which, by definition~\cite{Trautner:2025yxz}, act non-trivially on the kinetic terms.\footnote{In a very recent preprint~\cite{Grzadkowski:2026gkx}, it was suggested to define goofy trafos as combination of the exotic space-time transformation $x^\mu\mapsto\I x^\mu$ and a general C trafo. 
We refrain from such a definition since $(i)$ it does not cover all known cases of goofy trafos, e.g.\ it only applies to CP-type and not flavor-type goofy trafos (see~\cite{Trautner:2025yxz}), $(ii)$ in all cases known to us, it has never been necessary to include exotic space-time transformation to understand the physical consequences of goofy trafos, i.e.\ the associated RG fixed points.}
Hence, in the case of goofy Outs also the prefactors of the kinetic terms (i.e.\ the WFR coefficients) should be viewed as covariantly transforming couplings, implying that in this case also the evolution of WFR coefficients (anomalous dimensions) becomes non-perturbatively constrained.

\section{Example I}
As a first simple example, consider a QFT of a complex scalar field $\phi$ invariant under a $\Z{3}=\langle \mathsf{a}|\mathsf{a}^3=\mathrm{id}\rangle$ symmetry
that maps $\phi\mapsto\omega\phi$ and $\phi^*\mapsto\omega^2\phi^*$, where $\omega:=\mathrm{e}^{ 2\pi\I/3}$. $\phi$ and $\phi^*$ transform as $\rep{1'}$ and $\rep{1''}$ of $\Z{3}$, respectively. The most general renormalizable potential is ($m,\lambda\,\in\mathds{R}$, $\kappa\in\mathds{C}$)
\begin{align}\label{eq:potZ3}
	V=m^2 |\phi|^2+\kappa\,\phi^3+\kappa^*\phi^{*3}+\lambda|\phi|^4\;.
\end{align}
The group $\Z{3}$ has a $\Z{2}$ outer automorphism, $\mathrm{Out}(\Z{3})=\Z{2}$, that maps $\mathsf{a}\mapsto \mathsf{a}^2$ and $\rep{1'}\leftrightarrow\rep{1''}$. As an active transformation on the fields this Out corresponds to a map $\phi\leftrightarrow\phi^*$. Alternative to the active action on fields, this transformation can equivalently be described by a mapping in the parameter space of couplings, namely $\kappa\mapsto\kappa^*$, or in more detail $\mathrm{Im}\,\kappa\mapsto-\mathrm{Im}\,\kappa$. For a generic point in the parameter space, the theory is not invariant under this transformation.

The fact that an active transformation on the quantum fields exists that can be equivalently described by a mapping in the space of couplings, implies that the $\beta$ functions have to transform covariantly under it.
The only non-trivially transforming coupling is $\mathrm{Im}\,\kappa$, hence
\begin{equation}\label{eq:BetaImK}
\beta_{\mathrm{Im}\,\kappa}~\equiv~\mu\frac{\mathrm{d}\,\mathrm{Im}\,\kappa}{\mathrm{d}\mu}~=~\left(\mathrm{Im}\,\kappa\right)\times f_+(\lambda)\;,
\end{equation}
where $f_+$ denotes an arbitrary Out-invariant function. If the theory is required to be invariant under the Out, one has to request $\kappa=\kappa^*$, hence $\mathrm{Im}\,\kappa=0$, and, therefore, $\beta_{\mathrm{Im}\,\kappa}=0$. The \textit{existence} of the Out restricts the functional form of the $\beta$ function to \eqref{eq:BetaImK} to all orders in $\mathrm{Im}\,\kappa$,\footnote{%
Higher-order non-trivially transforming odd powers of $\mathrm{Im}\,\kappa$ are prohibited by dimensional analysis.
In fact, understanding scale transformations as Outs one can \textit{derive} dimensional power counting, which can give insights beyond naive dimensional analysis.\label{fot:ScaleOut}
}
while requiring the Out to be an actual symmetry explains the vanishing of the RG flow at the symmetry enhanced region $\mathrm{Im}\,\kappa=0$.
Physically, this Out corresponds to a CP transformation, and the region $\mathrm{Im}\,\kappa=0$ to a CP conserving theory.\footnote{
Of course, for this simple model even if $\kappa^*\neq\kappa$ we could always rephase $\phi$ (or equivalently include a non-trivial phase in the Out trafo) to show that physical CP is conserved irrespectively of the phase of $\kappa$.}

Next to the non-trivial $\Z{2}$ Out, there always exists the trivial Out. The trivial Out does not have to act trivially on the fields, specifically because the global phase of an Out is never constrained by the original group's algebra. Hence, there is an additional possible transformation $\phi\mapsto\e^{\I\alpha}\phi$, $\phi^*\mapsto\e^{\I\beta}\phi^*$. Taking $\alpha=\beta=\pi$ this can equivalently be viewed as a transformation $\kappa\mapsto-\kappa$, which constrains $\beta_\kappa=\kappa\,f_+(\lambda)$, hence, defines a preserved all-order stable RG fixed parameter region $\kappa=0$. For this imposed $\Z2$ transformation, the potential at renormalizable order, Eq.~\eqref{eq:potZ3}, has an accidental $\U{1}$ symmetry. The whole $\U{1}$ can be viewed as an enhancement by a continuous Out consistent with our arguments for the set of all possible choices of $\alpha=-\beta$.

\section{Example II}
\begin{figure}
    \centering
    \includegraphics[width=0.5\textwidth]{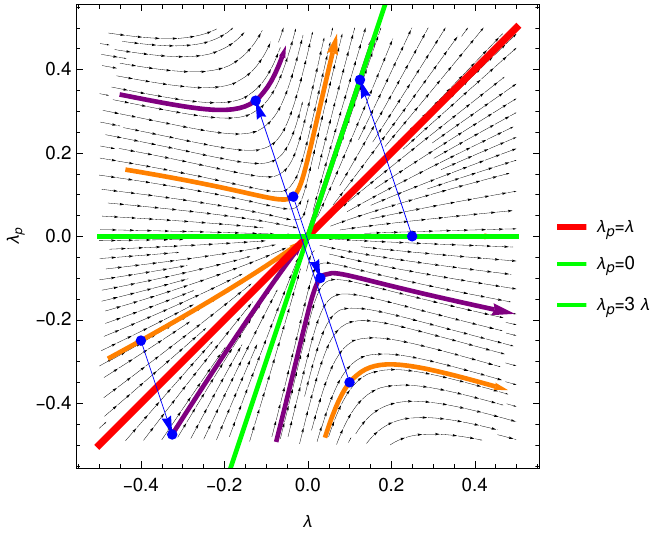}
    \caption{\label{fig:D8}
    Example II: RG flow in the $\lambda$-$\lambda_p$ plane. The red and green colored lines correspond to the (partial) fixed points listed in Eq.~\eqref{eq:D8FPs}.
    The blue arrows illustrate the action of the $\Z{2}$ outer automorphism transformation $u$, Eq.~\eqref{eq:D8Out}, as mapping on the parameter space and RG flow.}
\end{figure}

As second example, consider a QFT of fields $\phi_1$, $\phi_2\in\mathds{R}$ that form a doublet $\vec{\phi}\equiv(\phi_1,\phi_2)^\mathrm{T}$ invariant under the transformations
\begin{align}
&
\vec{\phi}
 \mapsto s\, 
\vec{\phi},\;
 t\,
\vec{\phi}
& 
&\text{with}&
&s:=\begin{pmatrix}
    1 & \\
    & -1 \\
   \end{pmatrix},\;
 t:=\begin{pmatrix}
    & 1\\
    1 & \\
   \end{pmatrix}.
&
\raisetag{18pt}
\end{align}
The invariant renormalizable potential is given by
\begin{equation}\label{eq:D8Pot}
 V(\phi_1,\phi_2)=m^2\left(\phi_1^2+\phi_2^2\right)+\lambda\left(\phi_1^4+\phi_2^4\right)+2\lambda_p\,\phi_1^2\phi_2^2\;.
\end{equation}
This model has been discussed in Sec.~II of~\cite{Berezhiani:2024rth} (and earlier in \cite{Iliopoulos:1980zd}). Based on an explicit computation of the $\beta$ functions at one loop, it was established there that the theory has three RG fixed parameter regions 
\begin{align}\label{eq:D8FPs}
&(i)\,\lambda=\lambda_p\;,& 
&(ii)\,\lambda_p=0\;,& 
&(iii)\,\lambda_p=3\lambda\;.&
\end{align}
The theory enjoys additional symmetries at these points that are referred to as ``emergent symmetries'' in~\cite{Berezhiani:2024rth}. We display the RG flow at two-loop order and the associated fixed lines in Fig.~\ref{fig:D8}. Let us show how to determine the symmetries and associated fixed point from Outs.

The group generated by $s$ and $t$ is $D_8$, the symmetry of the square, and $\vec{\phi}$ transforms in the $\rep{2}$ of $D_8$. This setup has an $\mathrm{Out}(D_8)=\Z{2}$ that corresponds to a permutation of generators $u(s)=t$, $u(t)=s$, acting on the doublet as 
\begin{align}\label{eq:D8Out}
 &\vec{\phi}\mapsto u\,\vec{\phi}\;,&
 &\text{with}&
&u=\frac{1}{\sqrt{2}}\begin{pmatrix}
    1 & \phantom{-}1\\
    1& -1 \\
   \end{pmatrix}.\;
&
\end{align}
This is \textit{not} a symmetry of the original QFT. However, the active transformation on the fields can \textit{equivalently} be described by a mapping in the parameter space
\begin{align}\label{eq:D8Outcouplings}
  &\begin{pmatrix}
  \lambda \\ \lambda_p
 \end{pmatrix}
 \xmapsto{u}
 \underbrace{
 \frac{1}{2}
 \begin{pmatrix}
 1 & \phantom{-}1 \\ 3 & -1
 \end{pmatrix}
 }_{=:\mathcal{O}_u}
 \underbrace{
 \begin{pmatrix}
   \lambda \\ \lambda_p
 \end{pmatrix}}_{\vec{\lambda}}.
\end{align}
We illustrate the action of this mapping on the RG flow by the blue arrows in Fig.~\ref{fig:D8}.

The fully coupled, non-perturbative system of $\beta$ functions, $\beta_{\lambda_i}\equiv\mu\mathrm{d}\lambda_i/\mathrm{d}\mu$, \textit{has to} obey constraint
\begin{align}\label{eq:D8BetaConstraintU}
\underbrace{
 \left.\begin{pmatrix}
  \beta_\lambda \\ \beta_{\lambda_p}
 \end{pmatrix}
 \right|_{\vec{\lambda}\mapsto\mathcal{O}_u\vec{\lambda}}
 }_{\substack{\beta=\beta(\lambda,\lambda_p)~\text{non-linear,} \\ \text{non-perturbative}}}
\stackrel{!}{=} 
\mathcal{O}_u
\underbrace{
\begin{pmatrix}
  \beta_\lambda \\ \beta_{\lambda_p}
\end{pmatrix}}_{=:\beta_{\vec{\lambda}}}
\;,
\end{align}
corresponding to Eq.~\eqref{eq:beta_trafo} for this special case. We have explicitly confirmed that this constraint is fulfilled using $\beta$ functions at three-loop order computed with PyR@TE~\cite{Sartore:2020gou, Poole:2019kcm} and RGBeta~\cite{Thomsen:2021ncy} (displayed in App.~\ref{app:D8}).

This constraint has to be fulfilled non-perturbatively and everywhere in the parameter space. Hence, one can turn this around to constrain $\beta$ function coefficients order by order. To see this, one can diagonalize the action of \eqref{eq:D8Outcouplings} by a reparametrization of couplings. The constraint~\eqref{eq:D8BetaConstraintU} then yields a factorization
\begin{equation}
 \beta_{\lambda-\lambda_p}=\left(\lambda-\lambda_p\right)\times f_+(\lambda,\lambda_p)\;,
\end{equation}
where $f_+$ here must be a $u$-invariant function,
\begin{equation}
f_+\left(\lambda,\lambda_p\right)~=~f_+\left(\frac12(\lambda+\lambda_p),\frac12(3\lambda-\lambda_p)\right)\;.
\end{equation}
This demonstrates our logic that the covariantly transforming $\beta_{\lambda-\lambda_p}$ can only be spanned by correctly transforming covariant combinations of parameters times arbitrary invariant functions. In the present case, the only correctly transforming covariant is $(\lambda-\lambda_p)$ such that the above factorization is the only possibility.

The Out $u$ becomes a symmetry of the theory exactly at the parameter region where all non-trivially transforming covariants of parameters vanish, i.e.\ here $\lambda=\lambda_p$. As a consequence of the above factorization, the system of $\beta$ functions loses a rank (i.e.\ the number of independent $\beta$ function is reduced).

This explains the origin of the RG fixed hyperplane $(i)\,\lambda=\lambda_p$. We note that the symmetry in parameter region $(i)$ is actually $\mathrm{O}(2)$~\cite{Berezhiani:2024rth} and, therefore, larger than just joining $s$, $t$ and $u$ transformations. This is an accidental symmetry of working at the renormalizable order and does not counteract our arguments. 

Next we discuss the symmetry enhanced parameter regions $(ii)$ and $(iii)$, which correspond to decoupling regions. Note that the two regions $(ii)$ and $(iii)$ are mapped onto each other under the transformation $u$, which shows that they 
describe exactly the same physics.\footnote{%
It is generally true that parameter regions related by an Out describe exactly the same physics. If there are multiple ``wedges'' of the parameter space related by Outs, then the physics described in each of the wedges is identical to the physics in all other wedges. This is discussed to some detail in~\cite{Fallbacher:2015rea,Trautner:2016ezn} and we come back to it below. To describe all possible measurements with parameters within a single given wedge may require a relabeling of external fields after learning about the results of measurements.}

\begin{figure*}[!ht!]
 \includegraphics[height=0.115\textwidth]{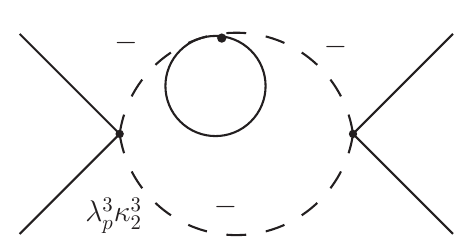}
 \includegraphics[height=0.11\textwidth]{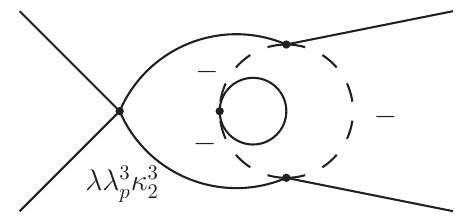}
 \includegraphics[height=0.11\textwidth]{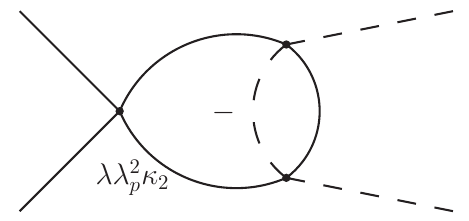}
 \includegraphics[height=0.11\textwidth]{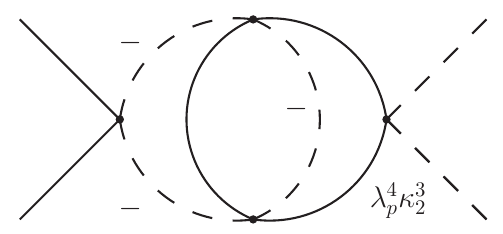}
\caption{\label{fig:loops}
Example diagrams for contributions to $\beta_\lambda$ (left) and $\beta_{\lambda_p}$ (right) at the two- and three-loop order. $\phi_1$ ($\phi_2$) propagators are drawn as solid (dashed) lines. We indicate the parametric order as well as the sign-flips of $\kappa_2$, corresponding to $\phi_2$-propagator sign-flips under the goofy transformation $w$.
}
\end{figure*}
It is more tricky to understand regions $(ii)$ and $(iii)$ from Outs, because we need to take into account \textit{goofy} transformations. To start this, note that the $D_8$-invariant potential \eqref{eq:D8Pot} (ex.\ mass terms) actually enjoys and additional invariance under the transformation\footnote{%
That goofy transformations transform real fields with imaginary numbers is a generic (but not necessarily a defining) feature, see the discussions in~\cite{Ferreira:2023dke, Haber:2025cbb, Ferreira:2025ate}.
}
\begin{align}
&\vec{\phi}\mapsto\,v\,\vec{\phi}&
&\text{with}&
v:=\begin{pmatrix} & -\I\\ \I&
\end{pmatrix}.
\end{align}
This transformation is $\textit{goofy}$, hence, by definition~\cite{Trautner:2025yxz}, acts non-trivially on the kinetic terms
\begin{align}
 K=Z_{ij}\,\partial_\mu\phi_i\partial^\mu\phi_j\cong\kappa_1(\partial_\mu\phi_1)^2+\kappa_2(\partial_\mu\phi_2)^2\;.
\end{align}
Here we have included wave function renormalization~(WFR) coefficients $Z_{ij}$, or $\kappa_{1,2}$ in a diagonal basis. All of these should be treated as couplings regarding their
explicit appearance in the $\beta$ functions as well as their
non-trivial transformation behavior under Outs. While one can always and w.l.o.g.\ work in a basis where the kinetic terms are canonical, we display the general form, and a diagonalized but not rescaled version here, to ease the following discussion. The transformation $v$ is a symmetry of the potential but explicitly broken by the kinetic terms, where it corresponds to a map $v:\,\kappa_{i=1,2}\mapsto-\kappa_i$.

This means that the potential is invariant under the full Pauli group $P=\langle t=\sigma_1,v=\sigma_2,s=\sigma_3\rangle$ which brings about a larger group of Outs. $\mathrm{Out}(P)\cong D_{12}$ is generated by $u$ and a new generator $w$ that acts as\footnote{%
In fact, $u$ and $w$ generate the group $\mathrm{SG}[192,963]$ (in GAP notation~\cite{GAP4}) which contains the Pauli group as a normal subgroup such that $D_{12}=\mathrm{Out}(P)=\mathrm{SG}[192,963]/P$. Closing only up to inner automorphisms is a common feature of explicit realizations of Outs, which are strictly speaking cosets of automorphisms.}
\begin{align}
&\vec{\phi}\mapsto\,w\,\vec{\phi}&
&\text{with}&
w:=\begin{pmatrix} 1 & \\ & \I
\end{pmatrix}.
\end{align}
The equivalent action on the parameters is given by
\begin{align}\label{eq:wOutcouplings}
  &
  \begin{pmatrix}
  \lambda \\ \lambda_p
 \end{pmatrix}
 \xmapsto{w}
 \begin{pmatrix}
 1 & \\ & -1
 \end{pmatrix}
  \begin{pmatrix}
   \lambda \\ \lambda_p
 \end{pmatrix}=:\mathcal{O}_w\,\vec{\lambda},&
\end{align}
and $\kappa_2\xmapsto{w}-\kappa_2$. The corresponding constraint on the $\beta$ functions is given by (again corresponding to Eq.~\eqref{eq:beta_trafo})
\begin{align}\label{eq:D8BetaConstraintW}
 &\mathcal{O}_w\,\beta_{\vec{\lambda}}
\stackrel{!}{=}
 \left.
 \beta_{\vec{\lambda}}(\lambda,\lambda_p,\kappa_1,\kappa_2)
 \right|_{\vec{\lambda}\mapsto\mathcal{O}_w\vec{\lambda},\,\kappa_2\mapsto-\kappa_2}\;.
 &
\end{align}
It is crucial to take into account the goofy nature of the transformation, here in the form of sign flips of $\kappa_2$, in \textit{all} $\beta$ functions, otherwise the relation is not fulfilled. 

For example, the $\kappa_2$ sign flip explains why $\beta_\lambda$ can have a cubic dependence on $\lambda_p^3$ proportional to $\lambda_p^3\kappa_2^3$, or why $\beta_{\lambda_p}$ can have a quadratic dependence proportional to $\lambda\lambda_p^2\kappa_2$.
None of these can be understood by considering simple manual sign flips of $\lambda_p$ while ignoring the sign flips of $\kappa_2$.
We have explicitly checked the constraint \eqref{eq:D8BetaConstraintW} to 3-loop level, manually taking into account internal propagator sign flips of $\phi_2$ diagram by diagram, see Fig.~\ref{fig:loops} for examples.
This is a workaround, necessary because all presently available RG codes assume to start in the canonical basis. This assumption is not wrong and can always be done, but it does not allow us to display the explicit dependence on the WFR $Z_{ij}$ here. Our workaround can be performed for any Out by considering the transformation in the corresponding ``Out-eigenbasis'', where the action on the kinetic term can at most amount to sign flips~\cite{Trautner:2025yxz}. Nonetheless, we strongly encourage a more general formulation of $\beta$ functions in the future, which should display the explicit the dependence on~$Z_{ij}$. 

Just like Eq.~\eqref{eq:D8BetaConstraintU}, also Eq.~\eqref{eq:D8BetaConstraintW} must be fulfilled irrespectively of any other assumption. Our general logic then implies that the non-trivially transforming $\beta_{\lambda_p}$ must be of the general form
\begin{equation}
\beta_{\lambda_p}=\lambda_p \times f_{+}(\lambda,\lambda_p,\kappa_2)+ \kappa_2\times g_{+}(\lambda,\lambda_p,\kappa_2)\;,
\end{equation}
spanned by the two covariantly transforming couplings $\lambda_p$ and $\kappa_2$, and two $w$-invariant functions $f_+$ and $g_+$.\footnote{The likewise correctly transforming covariant ``coupling'' coefficient $m_2^2$ does not enter $\beta_{\lambda_p}$ by dimensional analysis, which is nothing but our argument applied to scale transformations, see footnote~\ref{fot:ScaleOut}.}

It is easy to understand that $g_+$ must be proportional to $\lambda_p$, and as a $w$-even function in above decomposition can only depend on even powers of $\lambda_p$. 
This explains that in the parameter region $(ii)\,\lambda_p=0$, where transformation $w$ joins the preserved symmetry of the potential, one has $\beta_{\lambda_p}=0$ to all orders (even if $\kappa_2\neq0$) such that the system of $\beta$ functions loses a rank. This explains the all-order RG stability of parameter region $(ii)$ even though the goofy transformation is explicitly broken by the presence of the kinetic term $\kappa_2\neq 0$. This shows that the breaking by $\kappa_2$ here is soft, in the sense that it does not re-generate other goofy-non-invariant operators.

Lastly, also parameter region $(iii)\,\lambda_p=3\lambda$ can be understood by a goofy Out, corresponding to a transformation $z$, with action on the fields
\begin{align}
&\vec{\phi}\mapsto\,z\,\vec{\phi}&
&\text{with}&
z:=\frac12\begin{pmatrix}
    1+\I& 1-\I\\
    1-\I & 1+\I \\
\end{pmatrix}.
\end{align}
The alternative action on the couplings is given by
\begin{align}\label{eq:zOutcouplings}\raisetag{33pt}
  &
  \begin{pmatrix}
  \lambda \\ \lambda_p
 \end{pmatrix}
 \xmapsto{z}
 \underbrace{
 \frac12
 \begin{pmatrix}
 -1 & 1 \\ \phantom{-}3 & 1
 \end{pmatrix}
 }_{\mathcal{O}_z}
  \begin{pmatrix}
   \lambda \\ \lambda_p
 \end{pmatrix}
 &
 &\text{and}&
 &Z_{ij}\xmapsto{z}z^\mathrm{T}Z_{ij}z\;.&
\end{align}
This transformation is preserved for $\lambda_p=3\lambda$. We do not need to discuss $z$ separately, as it is already contained in the full Out group generated by $u$ and $w$ as $z=uwu^{-1}$. The action of $u$ corresponds to  the reflection highlighted by the blue arrows in Fig.~\ref{fig:D8}, hence, identifies the regions $(ii)$ and $(iii)$. Correspondingly, the constraint imposed on the $\beta$ functions by $z$ is automatically fulfilled since Eqs.~\eqref{eq:D8BetaConstraintU} and~\eqref{eq:D8BetaConstraintW} are already fulfilled. 

We stress again that taking into account the fully off-diagonal transformation of WFR $Z_{ij}$ in the $\beta$ functions requires a 
general formulation that displays this dependence, which is currently not available. For a trafo with general action on $Z_{ij}$, like $z$, 
one can always rotate to a basis where the action of $z$ on $\vec{\phi}$ is diagonal, and then use the above workaround of taking into account propagator sign flips. For $z$, this would look like $w$ in the original basis. Even though such a workaround is always possible, we reiterate our encouragement to provide a formulation of $\beta$ functions that takes into account the most general $Z_{ij}$-dependence, such that all-order constraints in the future can be imposed and checked in the most general basis.

Finally, we note that \textit{goofy} transformations unavoidably act non-trivially on the mass terms~\cite{Trautner:2025yxz, deBoer:2025jhc, Trautner:2025prm}. By contrast, for \textit{regular} Outs, the mass terms can, but do not have to be affected, as is the case for $u$. For example, imposing $w$ on the potential, the mass assignment $m_1\neq0$, $m_2=0$ is radiatively stable to all orders (explicitly checked up to two loops using RGBeta~\cite{Thomsen:2021ncy}). This is despite the fact that such mass assignment explicitly breaks $t$, and $\kappa_2$ breaks $w$, because both of these breakings are soft. The fact that a mass hierarchy can be radiatively stable may not be a surprise here, as the $w$-symmetric potential with $\lambda_p=0$ corresponds to a $\phi_1$-$\phi_2$-decoupling regime.\footnote{It is known that decoupling can be understood as symmetry enhancement in the form of dual Poincar\'e symmetry~\cite{Foot:2013hna}. Here we have shown that decoupling can also be understood by an enhanced goofy symmetry. This highlights once more the potential connection of goofy and space-time transformations which certainly deserves further attention.}
The goofy protection of mass hierarchies can work also in more general cases where not all portal terms vanish, see e.g.\ some of the 2HDM models in~\cite[Tab.~1]{Trautner:2025yxz} and the explicit BSM example in~\cite{deBoer:2025jhc}.

As a remark, we note that while constraints \eqref{eq:D8BetaConstraintU}, \eqref{eq:D8BetaConstraintW} (and similar for $z$) are very restrictive on the $\beta$ functions they do not entirely fix the functions. For the two existing independent Outs $u$ and $w$, here, the conditions imply $4$ constraints at \textit{each} order of a power series expansion ansatz of the most general possible $\beta$ functions. This is sufficient to warrant the exposed factorizations, but it does not fully fix the remaining invariant functions. We observe that the $\beta$ functions have to fulfill conditions that remind us very much of the defining constraints of modular forms. Indeed, $\mathcal{O}_{u,w,z}$ are all non-unitary, finite modular transformations. More research is necessary to clarify the ultimate power of the Out constraints.

\section{General Argument (in formulae)}\label{sec:GenArg2}
Lastly, we present our general arguments in formulae. A more detailed version of this discussion can be found in~\cite[Sec.~7]{deBoer:2025ncy}. An outer automorphism $u:g\to u(g)~\forall\,g\in G$ maps an irrep $\rep{r}$ of $G$ to an irrep $\rep{r'}$, determined by the consistency condition~\cite{Holthausen:2012dk,Fallbacher:2015rea,Trautner:2016ezn}
\begin{equation}\label{eq:consistency_condition}
\rho_{\rep{r}}(u(g))=U\rho_{\rep{r'}}(g)U^{-1}, \qquad\forall\,g\,\in\,G\;,
\end{equation}
where $\rho_{\rep{r^{(\prime)}}}(g)$ are matrix representations of $g$
and $U$ is the matrix representation of the Out.\footnote{%
Fixing the bases of $\rep{r}$ and $\rep{r'}$, the consistency condition determines $U$ up to a central element of the group and a global phase.
} 
This implies that $u$ maps a field $\phi_a$ in irrep $\rep{r}_{a}$ to a field $\phi'_a$ in irrep $\rep{r}_a^{\boldsymbol{\prime}}$ as
\begin{equation}\label{eq:out_field_trafo}
	\phi_a\,\xmapsto{u}\,U\phi'_a=:\tilde\phi_a=\mathcal{F}(\phi_a).
\end{equation}
Here $\mathcal{F}(\phi_a)$ should be understood as a function of \textit{all} present fields. The transformation $\phi\mapsto\tilde\phi$, by definition, is not a symmetry of the action.\footnote{Otherwise the Out would be part of the actual symmetry group, and one should study the Outs of that group.} 
However, the following statement is true.
\begin{theorem}\label{th:finding1}
If for every field $\phi_a$, also $\phi'_a$ is part of the QFT\,\footnote{This is equivalent to demanding that the field content of the model can be written in complete representations of the Out group. If this is not the case, the Out is \textit{maximally} broken by the field content of the QFT, see footnote~\ref{fot:maximalbreaking}.}, then the Lagrangian fulfills
\begin{equation}\label{eq:Lag_trafo}
	\mathcal{L}[\phi_a,\lambda_n]=\mathcal{L}[\tilde\phi_a,F_n(\lambda)]\;,
\end{equation}
for a set of functions $F_n(\lambda):=F_n(\{\lambda_m\})$ determined by the Out $u$, and $n$ denotes the independent couplings (incl. all possible WFR coefficients, masses, etc.).
\end{theorem}
This statement has been shown in Ref.~\cite{Fallbacher:2015rea}.  We repeat the arguments in App.~\ref{sec. proof lag trafo} (see also~\cite[Sec.~6.2f]{Trautner:2016ezn}).

\begin{corollary}\label{coll:sym_enhanced}
If the couplings fulfill  $F_n(\lambda)=\lambda_n$, then the transformation $\phi\mapsto\tilde\phi$ leaves the Lagrangian invariant. If this holds for a subgroup of $\Out{G}$ then this subgroup is a symmetry of the theory. 
\end{corollary}
\begin{corollary}\label{coll:fixed_hyperplane}
For every subgroup of \Out{G}, the fixed hypersurface defined by $F_n(\lambda)=\lambda_n$ corresponds to a region in parameter space with symmetry enhanced by this subgroup of \Out{G}. If this subgroup is non-anomalous, this also corresponds to an RG fixed hyperplane.
\end{corollary}

Both corollaries directly follow from Finding~\ref{th:finding1} and the usual intuition about 't\,Hooft's technical naturalness~\cite{tHooft:1979rat}.

\enlargethispage{0.5cm}
\begin{theorem}\label{th:finding2}
$\mathcal{L}[\phi_a,\lambda_n]$ and $\mathcal{L}[\phi_a,F_n(\lambda)]$ lead to the same physical predictions. Therefore, these parts of the parameter space are physically redundant.
\end{theorem}
Also this statement has been shown in Ref.~\cite{Fallbacher:2015rea} but we repeat the arguments here in a different form.
Consider the generating functional
\begin{equation}\label{eq:gen_funct}
	Z[J]=\int\mathcal{D}\phi_a\exp\left(i\int d^4x\ \mathcal{L}[\phi_a,\lambda_n]+J\phi_a\right)\;.
\end{equation}
By invariance under field redefinitions, 
\begin{equation}
	Z'[J]=\int\mathcal{D}\phi_a\exp\left(i\int d^4x\ \mathcal{L}[\phi_a,\lambda_n]+J\tilde\phi_a\right)
\end{equation}
gives exactly the same physical predictions as $Z[J]$.\footnote{By same physical predictions we mean the same S-matrix elements~\cite{Borchers:1960,*Chisholm:1961tha,*Kamefuchi:1961sb}. A modern derivation is given in~\cite{Manohar:2018aog}, see also e.g.~\cite{Kallosh:1972ap,*Passarino:2016saj,*Criado:2018sdb} and references therein.}
Using Eq.~(\ref{eq:Lag_trafo}) we can rewrite this to
\begin{align}\label{eq:gen_funct prime}\notag
\hspace{-0.2cm}Z'[J]=&\int\mathcal{D}\phi_a\exp\left(i\int d^4x\ \mathcal{L}[\tilde\phi_a,F_n(\lambda)]+J\tilde\phi_a\right)\\
=&\int\mathcal{D}\phi_a\exp\left(i\int d^4x\ \mathcal{L}[\phi_a,F_n(\lambda)]+J\phi_a\right). 
\end{align}
where we assumed $\det\left[\frac{\delta \mathcal{F}[\phi]}{\delta\phi}\right]=1$ in the final step, which is the requirement of anomaly freedom.

Since Eqs.~\eqref{eq:gen_funct} and~\eqref{eq:gen_funct prime} give the same physical predictions, the function $F_n(\lambda)$ maps parameters $\lambda$ to a physically redundant region of the parameter space.

\begin{theorem}
The redundancy in parameter space also manifests itself in the structure of the RG equations. Consider the $\beta$ functions as functions of the couplings as arguments (i.e. $\beta_{\lambda_n}(\,.\,)$), then Eq.~\eqref{eq:beta_trafo} holds.
\end{theorem}
To show this, note that the generating functionals in Eqs.~\eqref{eq:gen_funct} and~\eqref{eq:gen_funct prime} have the same functional form.
Hence, the $\beta$ functions in both theories have the same functional form. That is,
\begin{align}
    &\mu\frac{\mathrm{d}\,\lambda_n}{\mathrm{d}\mu}~=:~\beta_{\lambda_n}(\lambda_k)\;, \quad\text{and, therefore,}&\\\label{eq:betaflambda}
    &\mu\frac{\mathrm{d}\,F_n(\lambda)}{\mathrm{d}\mu}\equiv\beta_{F_n(\lambda)}(F_k(\lambda))=\beta_{\lambda_n}(F_k(\lambda))\;.&
\end{align}
The last equality is decisive here. It should be read as ``the beta function of the $n$-th operator'' in a decomposition like \eqref{eq:L_in_Op} must be the same (no matter what the couplings are) as the operators agree for both theories (of course, for the argument of the beta function the transformed couplings have to be used). Eq.~\eqref{eq:beta_trafo} then  follows by applying the chain rule to the far l.h.s. of Eq.~\eqref{eq:betaflambda}.

The RG stability of the Out-invariant hypersurface is then easy to see. Take $\lambda_n'\equiv F_n(\lambda)$, then
\begin{equation}
\begin{split}\hspace{-0.4cm}
\left.\left[\mu\frac{\mathrm{d}}{\mathrm{d}\mu}\left(\lambda'_n-\lambda_n\right)\right]\right|_{\lambda'=\lambda}=&~
\left.\beta_{\lambda'_n}(\lambda'_k)-\beta_{\lambda_n}(\lambda_k)~\right|_{\lambda'=\lambda} \\
\stackrel{(\ref{eq:betaflambda})}{=}\hspace{-0.1cm}&~ \left.\beta_{\lambda_n}(\lambda'_k)-\beta_{\lambda_n}(\lambda_k)~\right|_{\lambda'=\lambda}\\
=&~\,\beta_{\lambda_n}(\lambda_k)-\beta_{\lambda_n}(\lambda_k)=0.
\end{split}
\end{equation}
This establishes all of the points we had already explicitly demonstrated on our examples.

\section{Discussion}
We have used Outs to find explicitly broken possible realizable symmetries, and used this to formulate sufficient condition for their associated RG fixed points. Once known, our result may appear trivial and intuitive just like 't\,Hooft's technical naturalness, but we could not find our arguments  being pointed out in the literature before.

What about other fixed points? Is it not only sufficient but also necessary that there is an Out enhancement at every fixed point? One thing is for sure, scale symmetry is enhanced at any fixed point, by definition, corresponding to the vanishing of a $\beta$ function. Furthermore, scale symmetry \textit{is} an outer automorphism of the Lorentz and Poincar\'e groups~\cite{Buchbinder:2000cq}. Hence, it seems plausible that symmetry enhancement by an Out is not only sufficient but also necessary for a fixed point. 

This begs some hope, that using an analogous formulation of what we have used here, but for scale transformations (which, keep in mind, are generally anomalous), allows us to roll over the all-order action on covariantly transforming operators to all-order covariantly transformation of couplings. For scale (conformal) transformations, the according transformation on the couplings \textit{is} the RG flow.

Even though the dilatation transformation is anomalous
and our general argument of Finding~\ref{th:finding2} does not apply in this form, the dilations are Outs of the Poincar\'e group. Hence, it should be expected that their action on fields can equivalently be described by a mapping in the parameter space of the theory. Indeed, for the marginal couplings in our examples, a scale transformation\footnote{The action of the active scale transformation on space-time coordinates is $x_\mu\mapsto\e^{\sigma}x_\mu$, and for scalar fields $\phi(x)\mapsto\e^{\sigma}\phi(\e^{\sigma}x)$.}
with generator $\sigma$ corresponds to
\begin{equation}
    \lambda_n\mapsto \lambda_n+\sigma\beta_{\lambda_n}(\lambda)+\mathcal{O}(\sigma^2)\;.
\end{equation}
Since the functional form of the operators does not change, the analogue of Eq.~(\ref{eq:beta_trafo}) can be derived and to first order in $\sigma$ reads (sum over $m$) 
\begin{equation}
\begin{split}
\beta_{\lambda_n}
\left(\lambda_k+\sigma\beta_{\lambda_k}(\lambda)\right)&=
\beta_{\lambda_n}(\lambda)+\sigma\frac{\partial\,\beta_{\lambda_n}(\lambda)}{\partial \lambda_m}\beta_{\lambda_m}(\lambda).\\
\end{split}
\end{equation}
To leading order in $\sigma\ll 1$ this is trivially fulfilled, but the analogous constraint must hold to all orders. We have explicitly confirmed this for both of our examples at three loops up to order $O(\sigma^2\lambda^4)$ (or mixed quartic powers involving $\lambda_p$ in example~II). In general, the constraint gives rise to a nested and delayed differential equation that is not easy to solve. It should help to arrange contributions order-by-order in powers of $\sigma$ and make use of the fact that the constraint has to hold for any $\sigma$.

\enlargethispage{3cm}
The question is whether the constraints that can be derived from the action of a scale transformation Out are enough to non-perturbatively compute the $\beta$ function to all order,
similar to the all-order constraints we have derived from the Outs of the enhancement of the linear global symmetries. While the latter already explain multiple factorizations of the beta functions into Out co- and invariants, they do not completely constrain the invariantly transforming parts of the $\beta$ function.

For the examples discussed here, we found that the couplings and beta functions transform as representations of a finite modular group. This leads us to formulate the simple conjecture that closed form expression of the beta functions may have something to do with modular forms in general. 

Finally, we mention that our logic has so far worked without flaw for any model that we have considered. 
The oldest example we are aware of is the 3HDM with $\Delta(54)$ symmetry and $S_4$ Out~\cite{Fallbacher:2015rea}, where the covariant trafo of stationary points and $\beta$ functions under Outs has first been observed. A more recent example is the case of the 2HDM~\cite{Trautner:2025yxz}, where it was first understood that also \textit{goofy} transformations have to be included to cover all Outs. Some more examples are also found in~\cite{deBoer:2025ncy}. Our arguments work for the whole variety of models discussed in~\cite{Berezhiani:2024rth}. This means that neither do Outs have to be discrete for the logic to work, nor do they have to correspond to global symmetries but can also be local or space-time transformations.

Furthermore, note that in general it is true that if a consistent Out exists, one can extend a given model by that Out. For the resulting model (after imposing the Out as symmetry) the question about existing Outs has to be asked again. In this respect, note that in trees or chains of groups and subgroups, Outs can appear and disappear at all stages in a way that cannot be computed in a closed form for arbitrary groups. For finite groups this seems to be part of the group ``extension problem.''

A case that needs further investigation is the possibility of 
``unorthodox'' symmetry extensions~\cite{Doring:2024kdg}, which, by definition, do not correspond to symmetry enhancement by Outs. For example, this case arises by starting with a model that has a small symmetry group, which is then extended to a larger, simple group.\footnote{Simple groups, by definition do not have proper normal subgroups, hence, never can be constructed by a ``normal extension'' of a group using Outs.} The smallest finite simple group is the alternating group $A_5$, and one could, for example, start a model with a $\Z{3}\subset A_5$ subgroup, then impose the additional $\Z2$ generator. We expect that adding the necessary generator to arrive at $A_5$ enforces a non-invertible constraint on the parameters of the model, such that our discussion would need to be modified to take this into account. 

\section{Conclusions}
We have shown that the presence of an outer automorphism of a QFT provides a sufficient condition for the existence of a fixed hyperplane in the RG flow. This arises because couplings of the QFT transform covariantly under the outer automorphism, and this covariant transformation behavior is carried over to the beta functions of the theory. The system of beta functions automatically preserves the Out as a symmetry. This implies an all-order non-perturbative constraint on the beta functions which explains that the beta functions are spanned by Out-covariant combinations of parameters. At the Out-enhanced point in the parameter space, all non-trivially transforming covariant couplings vanish, which likewise explains the exact vanishing of some of the beta functions (reduction of rank of the system of beta functions), explaining the RG fixed point. In this way, the outer automorphism emerges as a symmetry of the theory at the fixed point. This is a generalization of 't ~Hooft's technical naturalness argument and explains the RG stability of symmetries in the first place.

Our argument is symmetry-based and can be applied to any QFT. It applies to couplings of all mass dimensions and does not require application of perturbation theory. Besides outer automorphisms (which are, in principle, computable from original symmetries of a QFT and its representation content), our argument applies to all possible transformations on fields and/or coordinates that can equivalently be described by a covariant transformation of couplings. In particular, our argument also applies to the recently discovered goofy transformations, and it is important to take those into account as to not miss any fixed points.

In the future it would be interesting to see how our argument carries over to non-linear transformation of couplings, and situations with anomalous symmetries. Most interesting will be a more detailed application to the conformal group and scaling (dilatation) transformations as outer automorphisms of the Poincar\'e group. We can presently not excluded that (at least for renormalizable theories) this may allow to compute the beta functions and anomalous dimensions to all orders in closed form.  

Beyond various applications in particle physics, our work motivates rethinking the whole program of renormalization, treatment of critical phenomena, phase transitions \textit{etc.}, always keeping in mind the power of outer automorphisms.

\pagebreak
\section*{Acknowledgments}
\vspace*{-0.7cm}\enlargethispage{1cm}
We thank \'Alvaro Pastor Guti\'errez for participating in an intermediate stage of this work.
A.T.\ thanks Simone Blasi for earlier discussions on the 3HDM beta functions, Howard E.\ Haber for discussion on goofy transformations,
as well as Michael Ratz for coining the phrase ``the symmetry of the beta functions is larger than the symmetry of the action.'' \\
The work of A.T.\ is supported by the Portuguese Funda\c{c}\~ao para a Ci\^encia e a Tecnologia (FCT) through projects \href{https://doi.org/10.54499/2023.06787.CEECIND/CP2830/CT0005}{2023.06787.CEECIND}, \href{https://doi.org/10.54499/UID/00777/2025}{UID/00777/2025}, and contract \href{https://doi.org/10.54499/2024.01362.CERN}{2024.01362.CERN}, partially funded through POCTI (FEDER), COMPETE, QREN, PRR, and the EU.

\bibliographystyle{utphys}
\bibliography{bib}

\appendix
\onecolumngrid
\section{\texorpdfstring{$\boldsymbol{\beta}$}{Beta} functions for Example II}\label{app:D8}\noindent
We used PyR@TE~\cite{Sartore:2020gou, Poole:2019kcm} and RGBeta~\cite{Thomsen:2021ncy} to derive the $\beta$ functions. Using the generic form
\begin{equation}
 \beta_{\lambda_n}(\lambda)=\sum_\ell \frac{\beta^{(\ell)}_{\lambda_n}(\lambda)}{(4\pi)^{2\ell}}\;,
\end{equation}
up to the two loop order we have 
\begin{align}
 &\beta_m^{(1)}/m= 24\lambda +8\lambda_p\;,&
 &\beta_m^{(2)}/m=-480 \lambda^2-160\lambda_p^2\;.&
 \end{align}
And up to the three loop order
\begin{align}
&\beta_\lambda^{(1)}/8= 9\lambda^2+ \lambda_p^2\;, \qquad\qquad \beta_\lambda^{(2)}/64= -51 \lambda^3 - 5 \lambda\lambda_p^2 - 4 \lambda_p^3\;,\\
&\beta_\lambda^{(3)}/64= \left(2592\,\zeta_3 + 3915\right)\lambda ^4+\left(384\,\zeta_3 + 808\right) \lambda  \lambda_p^3+\left(96 \zeta_3 + 51\right) \lambda_p^4+162\, \lambda ^2 \lambda_p^2\;,
\end{align}
as well as 
\begin{align}
&\beta_{\lambda_p}^{(1)}/16= \lambda_p \left(3 \lambda + 2 \lambda_p\right)\;, \qquad\qquad 
\beta_{\lambda_p}^{(2)}/192=- \lambda_p \left(5\lambda^2+12 \lambda  \lambda_p+3 \lambda_p^2\right) \;,\\
&\beta_{\lambda_p}^{(3)}/256=
\lambda_p \left[\left(432\,\zeta_3 +315\right)\lambda^2 \lambda_p+\left(288\,\zeta_3 +450\right)\lambda  \lambda_p^2+\left(48\,\zeta_3 +91 \right)\lambda_p^3+378 \lambda ^3\right]\;.
\end{align}

\section{Motivating \texorpdfstring{Eq.~\eqref{eq:Lag_trafo}}{Eq.(23)}}\label{sec. proof lag trafo}
Assume we have an action $S=\int d^4x\,\mathcal{L}$ invariant under some global symmetry group $G$ with fields $\phi_a$ transforming in some representations $\rep{r}{_a}$ of $G$ with matrix representation $\rho_{\rep{r}_i}(g)$, i.e.\ $\forall\,g\in G$
\begin{equation}\label{eq:symm_trafo}
	\phi_a\mapsto\rho_{\rep{r}_{a}}(g)\phi_a\;.\qquad\text{(no sum)}
\end{equation}	
One can write the Lagrangian in terms of $G$-invariant operators $\mathcal{O}_n[\phi_a]$ and associated independent couplings $\lambda_n$,
\begin{equation}\label{eq:L_in_Op}
	\mathcal{L}[\phi_a,\lambda_n]=\sum_n \lambda_n \mathcal{O}_n[\phi_a]\;.
\end{equation}
This sum shall include all possible independent operators consistent with all symmetries. For a general field redefinition $\tilde\phi_a:=\mathcal{F}(\phi_a)$, consider
\begin{equation}\label{eq:Ltilde_in_Op}
	\mathcal{L}[\phi_a,\lambda_n]=\mathcal{L}[\mathcal{F}^{-1}(\tilde\phi_a),\lambda_n]=:\tilde{\mathcal{L}}[\tilde\phi_a,\lambda_n]=\sum_k \tilde F_k(\lambda_n)\tilde{\mathcal{O}}_k[\tilde\phi_a]\;.
\end{equation}	
In the last step we decomposed $\tilde{\mathcal{L}}$ in a similar way to $\mathcal{L}$ before. Crucially, now only for Outs, we have $\tilde\phi_a=U\phi'_a$, implying that under symmetry transformation, Eq.~\eqref{eq:symm_trafo}, the fields $\tilde\phi_a$ transforms as
\begin{equation}
    \tilde\phi_a=U\phi'_a\mapsto U\rho_{\rep{r}'_a}(g)\phi'_a\stackrel{\mathrm{Eq.}\eqref{eq:consistency_condition}}{=}\rho_{\rep{r}_a}(u(g))U\phi'_a=\rho_{\rep{r}_a}(u(g))\,\tilde\phi_a\;,\qquad \forall\,g\in G\;,
\end{equation}
i.e.\ in the $\rho_{\rep{r}_a}$ representations of a different group element. Since this has to hold for all group elements, the invariant operators in the decomposition of $\tilde{\mathcal{L}}$ are constrained in exactly the same way as the invariant operators in the decomposition of $\mathcal{L}$.
Hence, for any operator $\tilde{\mathcal{O}}_{k}$ in \eqref{eq:Ltilde_in_Op} there must be an operator $\mathcal{O}_{n(k)}$ in \eqref{eq:L_in_Op} such that (technically this can be a linear combination of operators but the argument still holds)
\begin{equation}
	\mathcal{O}_{n(k)}[\tilde\phi_a]=\tilde{\mathcal{O}}_k[\tilde\phi_a].
\end{equation}
Therefore, we can write 
\begin{equation}
	\tilde{\mathcal{L}}[\tilde\phi_a,\lambda_n]=\sum_k \tilde F_k(\lambda_n)\tilde{\mathcal{O}}_k[\tilde\phi_a]=\sum_k\tilde F_k(\lambda_n)\mathcal{O}_{n(k)}[\tilde\phi_a]
	=:\sum_n F_n(\lambda_n) \mathcal{O}_n[\tilde\phi_a]=\mathcal{L}[\tilde\phi_a,F_n(\lambda_n)].
\end{equation}
Together with Eq.~\eqref{eq:Ltilde_in_Op} this implies Eq.~\eqref{eq:Lag_trafo}.
\end{document}